\documentclass[11pt]{article}
\usepackage{amssymb}\usepackage{amsmath}
\def\iindex{\index}
\usepackage{epsfig} 

\newtheorem{theorem}{Theorem}

\newtheorem{remark}{Remark}

\def\a{\alpha}
\def\d{\delta}

\def\sign{{\rm  sign\,}}

\def\const{{\rm const\,}}

\def\R{{\bf R}}

\def\b{\beta}

\def\X{{\cal X}}
\def\t{\theta}
\def\oo{\bar}

\newcommand{\be}{\begin{equation}}
\newcommand{\ee}{\end{equation}}
\newcommand{\bd}{\begin{displaymath}}
\newcommand{\ed}{\end{displaymath}}
\newcommand{\ba}{\begin{array}{ll}}
\newcommand{\ea}{\end{array}}
\newcommand{\baa}{\begin{eqnarray}}
\newcommand{\eaa}{\end{eqnarray}}
\newcommand{\baaa}{\begin{eqnarray*}}
\newcommand{\eaaa}{\end{eqnarray*}}

\def\oo{\bar}

\def\k{{\kappa}}

\def\T{{\cal T}}

\def\X{{\cal X}}

\def\x{x}
\date{ Submitted: December 12, 2013. Revised: January 29, 2014 } 
\title{Transmission of a continuous signal via one-bit capacity channel
\footnote{ accepted to {\em IEEE Signal Processing Letters}}}
\author{ Nikolai Dokuchaev}
      
\begin{document}
\def\break{}%
\def\brea{}
\def\breakk{}
 \maketitle
\begin{abstract}
We study the problem of  the transmission  of  currently observed
 time variable signals   via a  channel that is capable of
sending a single binary signal only for each measurement of the
underlying process. For encoding and decoding, we suggest a modification of the adaptive delta modulation algorithm. This modification ensures tracking of time variable signals. We  obtained  upper estimates for
the error for the case of noiseless transmission.
\end{abstract}
\let\thefootnote\relax\footnote{The author is with  Department of
Mathematics and Statistics, Curtin University, GPO Box U1987, Perth,
Western Australia, 6845 (email N.Dokuchaev@curtin.edu.au).This work  was supported by ARC grant of Australia DP120100928 to the author.}
\section{Introduction}
We study the problem of transmission of a currently observed
continuous time signal via a noiseless binary channel. The evolution
law for the underlying  continuous time signal is not supposed to be
known; only some mild conditions on the signal regularity are
imposed. In particular, the signal is not necessarily continuous,
and unexpected jumps may occur. We consider the situation where the
channel capacity is insufficient to send  in real time sufficiently accurate
approximations of the current measurements. Therefore,
the observed measurements have to be encoded, transmitted in the encoded
form, and decoded.  This problem may arise, for example,  for remote
control of underwater vehicles, since communication  is severely
limited underwater (see \cite{SB}). \index{Stitwell and Bishop \cite{SB}).}
\par
The paper suggests a
modification of the systems from \cite{WB,NE,DS}, \index{Wong and Brockett \cite{WB}, Nair
and Evans \cite{NE}, and Dokuchaev and Savkin \cite{DS},} where
limited capacity digital channels  were studied
 in stochastic setting. In   \index{Wong and Brockett} \cite{WB},  a related filtering problem was considered for
the case of bounded random disturbances. A case of decreasing
Gaussian disturbances was studied in \index{Nair and Evans} \cite{NE} for a
scalar system. In \index{Dokuchaev and Savkin} \cite{DS}, a filtering
problem was studied for the case of non-decreasing Gaussian
disturbances for vector processes. The present paper considers an extreme case of a  binary  channel, i.e., one-bit capacity channel that can
transmit a single binary signal for a single measurement of the
underlying process.  This channel connects  two subsystems of a
dynamical system. The first subsystem, that is called Encoder,
receives the real-valued measurements and converts them into a
binary symbolic sequence which is sent over the communication
channel. For each measurement, only one single bit signal can be
sent. The second subsystem (Decoder) receives this symbolic sequence
and converts it into a real-valued state estimate.
Therefore, the effectiveness of the entire system is defined by the effectiveness
of the binary encoding algorithm. This encoding problem was widely studied in the literature.
In non-causal setting, some important results can be found in \cite{CD,KIR,M,KP}; see also the bibliography therein.
In \cite{CD,KIR,KP}, the encoding was studied in the framework of the  sampling theory and Fourier analysis.
In \cite{M}, a sequential binary estimator based on stochastic contamination
was obtained   for continuous processes.
 \par
For our particular task, we have restrictions of causality. To satisfy this condition, we suggest  a modification of the adaptive delta modulation
algorithm  introduced  by Jayant \cite{Ja}   for voice transmission; see more recent development
in  \cite{DDA} and the bibliography therein.  The suggested algorithm
ensures stable  tracking of  time variable signals using just one bit for each measurement. The
algorithm does not depend on the parameters of the evolution law and
the distributions of the underlying process.  We obtained the upper estimates
of the error for the case of a
noiseless transmission.

\index{\par The remainder of this paper proceeds
as follows. In Section 2, we introduce the class of systems under
consideration formulate the main results.
  Section 3 presents an illustrative example. Section 4
contains brief discussion and suggest future developments.
The proofs of all the results are given in Appendix.}

\section{Problem statement and the  result}
Let $\x(t)$ be a  continuous time state process observed at times $t_k=k\d$, $k=0,1,2,...$, where  $\d>0$ is given.
\par
Suppose estimates of the current state $\x(t)$  are required at a distant
location, and are to be transmitted via a digital communication channel
such that only one bit of data may be sent at each time $t_k$, i.e., a binary channel.
For this task,  we consider a system which consists of the encoder, the transmission
channel, and the decoder.  For each observation  $x(t_k)$,
the encoder produces a one-bit symbol $h_k$ which is transmitted
via the channel and then received by the decoder; the decoder
produces an estimate $y(t)|_{[0,t_k]}$ which depends only on $h_1,....,
h_k$.
 In other words, the process $\x(t)$
is supposed to be sampled at times $t_k$, encoded, transmitted via the channel and then
decoded. We assume that the transmission is noiseless.
\iindex{  The block diagram of this system
is shown in Figure \ref{block1}.
\begin{figure}[htpb]
\begin{center}
\setlength{\unitlength}{0.7mm}%
\begin{picture}(50,10)(50,20)
\put(-24,19){\vector(1,0){22}} \put(-18,22){\sf  $x(t)$}
\put(-2,11){\framebox(30,14)} \put(0,17){\sf sampling}
\put(28,19){\vector(1,0){36}} \put(40,23){$x(t_k)$}
\put(64,11){\framebox(30,14)} \put(73,17){\sf Encoder}
\put(94,19){\vector(1,0){42}} \put(100,23){$h_k\in\{-1,1\}$}
\put(106,12){\sf Channel} \put(136,11){\framebox(27,14)}
\put(138,17){\sf Decoder} \put(163,19){\vector(1,0){20}}
\put(167,23){$y(t)$ }
\end{picture}
\end{center}
\caption{\sf Block
diagram of the estimator.}
\label{block1}
\end{figure} }
\par
It is important that, for each sample  for each sampling point $t_k$, only one bit of information can be transmitted.
The corresponding algorithm is suggested below.


Let real numbers $y_0$,  $M_0>0$, $\oo M>0$, and  $a\in (1,2]$ be given
parameters that are known both to the encoder and the decoder. The algorithm
can be described as follows.
\begin{enumerate}
\item Sample values $\x(t_k)$ are taken;
\item The  encoder computes  a sequence $\{(y_k,M_k)\}_{k\ge 1}\subset \R^2$ and produces  a sequence of binary symbols $\{h_k\}$
consequently for $k=-1,0,1,2,...$ by the following rule: $h_{-1}=1$, and
\baa \label{y}
&& h_k=\left\{
         \begin{array}{ll}
           1, & \hbox{if} \quad  y_{k}<\x(t_k) \\
           -1, & \hbox{if} \quad  y_{k}> \x(t_k)\\
          -h_{k-1}, & \hbox{if} \quad  y_{k}=\x(t_k),\\\end{array}
       \right.\eaa
where
\baa&& y_{k}=y_{k-1}+h_{k-1}M_{k-1}\d,\quad k=1,2,...\label{yy}\eaa
\baa
\hphantom{}  M_k=\left\{
         \begin{array}{l}
           aM_{k-1},\quad  \hbox{if  $k\notin\T$ and $k-1\notin\T$}  \\
           M_{k-1},\quad  \hbox{if $k\notin\T$ and $k-1\in\T$}  \\
           \max(a^{-1}M_{k-1},\oo M   ),\quad \hbox{if} \quad k\in\T,
         \end{array}    \right.\label{M}
\eaa
and where $\T=\{k\ge 1: h_{k-1}h_k<0\}$.

\item The binary symbol $h_k$ is transmitted via the channel.
\item  The decoder computes the same  sequence $\{(y_k,M_k)\}_{k\ge 1}\subset \R^2$ using the received  values  $\{h_k\}$
by the same rule as the encoder.
\item
 Finally, the
decoder computes estimate $y(t)$ of the process $\x(t)$ as \baa
 y(t)=y_k+h_kM_k(t-t_k),\quad
t\in[t_k,t_{k+1}],\label{wx}
  \eaa
 \end{enumerate}
and where  $k=0,1,2,...$, $y(0)=y_0$.

Note that this algorithm represents  a modification of  the Jayant's adaptive delta modulation algorithm \cite{Ja,DDA}, where
it was assumed that, in our notations,
\baa
\oo M=0,\qquad M_k=aM_{k-1}\quad\hbox{if}\quad k-1\in\T\label{Ja}.
\eaa
 The novelty of algorithm (\ref{y})--(\ref{wx}) is that it allows three possible values of $M_k/M_{k-1}$ on each step instead of two, and uses two bit memory instead of one.
It was shown in \cite{DDA} that Jayant's algorithm allows to track benchmark constant processes.
  We show below that he suggested changes ensure stable tracking processes for variable in time underlying processes.
   In addition, we were able to estimate the tracking error.
\par
 Let $D>0$ be given.
 \par
 We consider continuous time processes $x(t)$ 
with a polynomial rate of growth, i.e., such that there exists $C>0$ and $c>0$ such that
\baaa |x(t+\t)|\le C(|x(t)|+\t^c),\quad t>0,\quad \t>0.\label{Cc}\eaaa
In addition, we assume that there are intervals $[\a,\b]\subset[0,+\infty)$ such that
     \baa
   \sup_{t\in[t_k,t_{k+1}]}|\x(t)-\x(t_{k})|\le D\d
    \label{D}\eaa for all for all $k$ such that $[t_k,t_{k+1}]\subset[\a,\b]$.  \index{there exists a sequence of real intervals $\{[a_m,b_m]\}_{m=0}^\infty$ such that
    $0< a_0<b_0\le a_1<b_1\le a_2<b_2\le ...$,   and \baa
   \sup_{t\in[t_k,t_{k+1}]}|\x(t)-\x(t_{k})|\le D\d.
    \label{D}\eaa for all $m\ge 0$, for all $k$ such that $[t_k,t_{k+1}]\subset[a_m,b_{m}]$. }
     \index{ \baa
    \sup_{m\ge 0}\,\,\sup_{k:\, [t_k,t_{k+1}]\subset[a_m,b_{m}]}\,
   \sup_{t\in[t_k,t_{k+1}]}|\x(t)-\x(t_{k})|\le D\d.
    \label{D}\eaa  there exists a sequence $\{q_k\}_{k=0}^\infty$ such that
    $0=q_0< q_1< ...$ and \baa
    \sup_{[t_k,t_{k+1}]\subset[q_k,q_{k+1}]}\,
   \sup_{t\in[t_k,t_{k+1}]}|\x(t)-\x(t_{k})|\le D\d.
    \label{D}\eaa}
In fact, $y(t)$ will be approaching $x(t)$ during these time intervals only. Therefore, a  good approximation is not feasible if these intervals
are too small or have too large gaps between them.
 \begin{remark} Clearly, (\ref{D}) holds if
  $x|_{[\a,\b]}$ is absolutely continuous and $|dx(t)/dt|\le D$. However,
 we prefer to use condition  (\ref{D}) since it  is less restrictive; in particular, (\ref{D}) holds for some  discontinuous on $[\a,\b]$  processes $x$.
   \end{remark}
\par

\begin{theorem}
\label{ThM} \begin{itemize} \item[(i)] Let
 $\tau=\inf\{m\in\T\}$. Then
\baa
\tau\le\inf\{m\ge 0:\ M_0(1+a+a^2+\cdots+a^m)\d\brea\ge
|y_0-\x(0)|+C(1+m^c\d^c)\,\}. \label{tau}\eaa
Here $C$ and $c$ are the constants from (\ref{Cc}). \item[(ii)]
Assume that $\oo M\ge 2D$. Let  $\t=3\log_a(M_{\tau}/\oo M)+6$. In this case, if (\ref{D}) holds for all $k\in\{\tau,\tau+1,\tau+2,...,\tau+\t\}$, then  there exists
an integer $\eta\in\{\tau,\tau+1,\tau+2,...,\tau+\t\}$ such that \baa
 M_\eta=
\oo M,\qquad |\x(t_\eta)-y(t_\eta)|\le (a\oo M+D)\d.\label{eta}\eaa
\item[(iii)]
Assume that $\oo M\ge 2D$, assume that (\ref{eta}) holds for some
integer $\eta\ge 0$, and assume that (\ref{D}) holds for $k\ge \eta$. Then  $M_k\in\{\oo M,a\oo M\}$ for all $k\ge\eta$ and $M_k=\oo M$
for all $k\ge \eta$ such that $k\in\T$.
 In addition, \baa |\x(t_k)-y(t_k)|\le (a\oo M+D)\d,\nonumber
\\
\sup_{t\in[t_k,t_{k+1}]}|\x(t)-y(t)|\le (a\oo
M+2D)\d, \label{1d} \eaa
for all  $k\ge \eta$.
 \end{itemize}
\end{theorem}
\par
{ The proof of the theorem  is given
in Appendix.}
\par
Let us discuss some implications of  Theorem \ref{ThM}. As can be seen,   $y(t)$
stats to approximate $\x(t)$ after the time $t_\eta$ and until (\ref{D}) is overstepped.
The time period $[0,t_\eta]$ is used to bring the value  $y(t)$  to a close proximity of $\x(t)$. The time period $[t_\tau,t_\eta]$ is used to reduce the value  $M_k$ from $M_\tau$ down to $\oo M$. The approximation error  can be significant during the time interval
$[0,t_\eta]$, if the distance $|y_0-\x(0)|$ is large.

If a jump of $x$ occurs at time $s>t_\eta$, then Theorem \ref{ThM} can be applied again for the initial time $t=s$ instead of $t=0$, and for $y_0$ and $M_0$ replaced
by $y(\k\d)$ and $M_{\k}$, where $\k=\min\{k: k\d\ge s\}$.

If $\oo M\ge 2D$, then, by Theorem \ref{ThM}, the process  $y(t)$ oscillates about the underlying process, and the error does not vanish even for constant $\x(t)$.
 On the other hand, a choice of small  $\oo M$ may lead to a larger time of proximity recovery after a jump of $\x$. \index{For example, for a constant process $\x(t)$ and $\oo M=0$,  $M_k$ converges to zero while  $x(t)$ is constant, and the approximation error is vanishing. However, if a jump of $\x$ occurs while $M_s$ is small, it will take longer  time to
 increase the size of  $M_s$  to be able to restore the proximity to  $\x$.}
\begin{remark} The  suggested algorithm is robust with respect to the errors caused by missed  or misread
signals $h_k$ for the models where the decoder is always
aware that a signal was missed or  misread, i.e., for the case of
the {\em so-called binary erasure channel}. \index{In particular,
this is true if the times $t_k$ are known and counted.} Obviously,
there are models of channels with noise where  these conditions are
not satisfied.   It could be interesting to find a way to modify an
algorithm such that it will be robust with respect to the
transmission errors for these models.
\end{remark}
\
\section{Illustrative examples}
In  numerical experiments, we compared the performance  of the Jayant's encoding algorithm \cite{Ja}  and modified version  (\ref{y})--(\ref{wx}). We observed that,  in all our experiments, the modified
version (\ref{y})--(\ref{wx}) ensures faster recovering of the proximity after a  jump of the underlying process. This is illustrated in Figures \ref{fig0}-\ref{fig1} presenting
the results of the applications of the  Jayant's encoding algorithm \cite{Ja} and  the suggested algorithm (\ref{y})--(\ref{wx})  of  a discontinuous piecewise continuous
process $x(t)$.
Figures \ref{fig0}-\ref{fig1} show the
process $x(t)$ and the
corresponding processes $y(t)$ for   $\d=t_{k+1}-t_k=0.04$ and for  $\d=t_{k+1}-t_k=0.02$ respectively, $t\in[0,2]$. With these sampling rates, transmission of the encoded
signals for $t\in[0,2]$ requires to transmit 50 bits only for $\d=0.04$
and 100 bits only for $\d=0.02$.  For these examples, we used $y_0=5$, $a=1.5$, and $M_0=2\d=2D$. The algorithm (\ref{y})--(\ref{wx}) was applied  with $\oo M=2\d$.
We used MATLAB for these calculations.
\setcounter{figure}{0}
\begin{figure}[ht]
\caption[]{Example of  a discontinuous input $x(t)$ and the
corresponding estimate $y(t)$ for $\d=0.04$ for Jayant's encoding algorithm \cite{Ja} and for the suggested algorithm (\ref{y})--(\ref{wx}).
}
\centerline{\psfig{figure=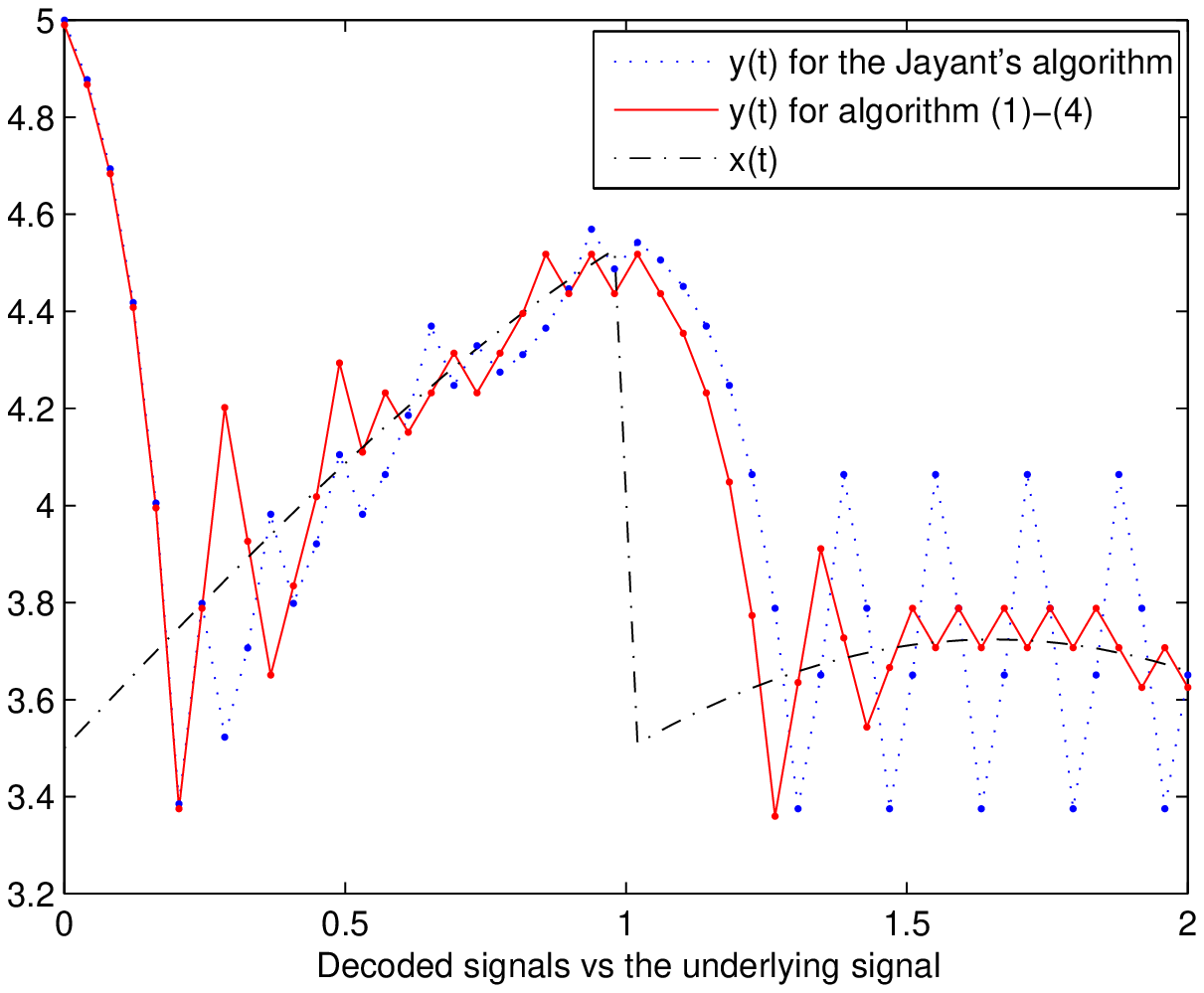,height=5.5cm}}
 \label{fig0}
\end{figure}
\begin{figure}[ht]
\caption[]{Example of  a discontinuous input $x(t)$ and the
corresponding estimate $y(t)$ for
$\d=0.02$, for Jayant's encoding algorithm \cite{Ja} and for the suggested algorithm (\ref{y})--(\ref{wx}).
}
\centerline{\psfig{figure=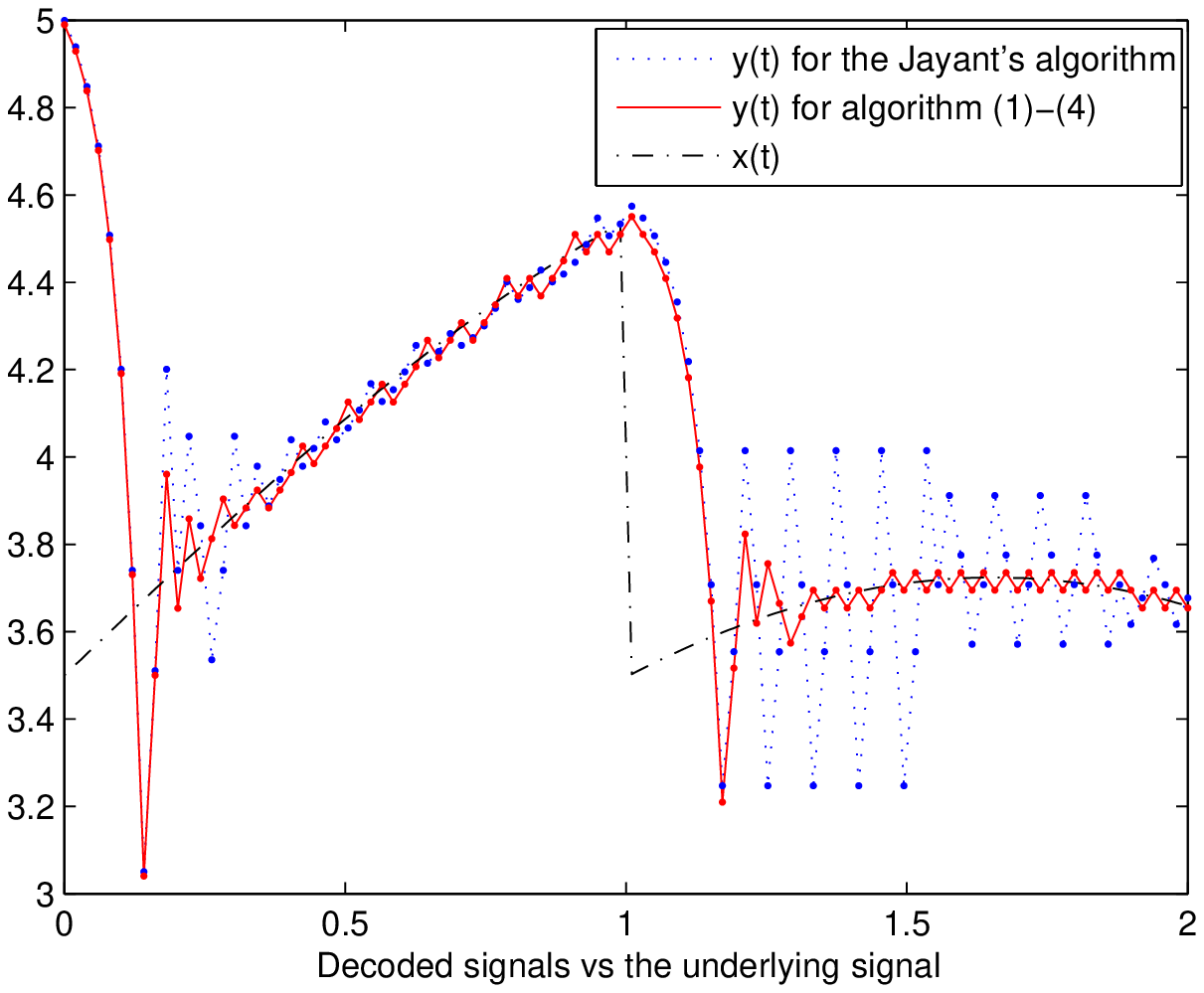,height=5.5cm}}
 \label{fig1}
\end{figure}
\section{Discussion and future developments}
\begin{enumerate}
\item The estimates in Theorem \ref{ThM} represent the upper bounds for the worst case scenario;
in practice, one should expect a better performance. A more
informative  estimate of the algorithm performance could be obtained
in the stochastic setting, such as the mean square error given
certain probabilistic characteristics of the input. In this setting,
optimal selection of the parameter $a$ could be investigated.
\index{
\\...
\\ It could interesting  to estimate the algorithm performance in the stochastic setting, such as
mean square error given certain probabilistic characteristics of the input. }
\item  It could be interesting to extend the algorithm on vector processes  $\x(t)$.
\item The presented algorithm  is causal, i.e., it collects
current information and does not require the future values of the
process. It could be interesting to estimate the loss of
the effectiveness caused by the causality restrictions in comparison with
the algorithms known in the rate-distortion theory   in non-causal setting, where an entire signal $x(t)|_{[0,T]}$ is known
and has to be encoded, for some given interval
$[0,T]$ (see, e.g., \cite{Cover}, Ch. 13).
\item  In theory, an arbitrarily close causal approximation in $L_2$-norm can be achieved
by binomial processes with a fixed rate of change for general stochastic square integrable
processes, including Ito  and jump processes \cite{D14}. However,
an algorithm of this approximation was not obtained therein. It could be
interesting to investigate if  the algorithm from the present paper
can be used to achieve this kind of  approximation.
\index{In this case, a continuous time function can be encoded via a
sequence of binary symbols, with the rate of one binary symbol for
each sampling point, rather than a real number for each sampling
point. This may provide a representation that requires less bits
than a representation via truncated Fourier series or splines. We
leave it for the future research.}
\end{enumerate}
\section*{Appendix: Proof of Theorem \ref{ThM}}
To prove  statement (i), it suffices to observe that \baaa
&&\inf_{k\le m}|y(t_k)-\x(t_k)|\le
|y_0-\x(0)|\breakk+C(1+m^c\d^c)-M_0(1+a+a^2+\cdots+a^m)\d. \eaaa
\par
Let us prove statement (ii).
 For integer numbers $s\ge 0$, set   \baaa
\quad
\tau(0)=\tau=\inf\{m\in\T\},\quad \tau(s)=\inf\{m> s:\  m\in\T\},\brea \quad s>0.
\eaaa Let us prove first that, for any
$s\in\T$, \baa\tau(s)\le s+3\label{tree}\eaa and \baa
 \sup_{s\le k\le \tau(s)}|\x(t_k)-y(t_k)|\le (M_{s-1}+D)\d, \nonumber\\
\sup_{s\le k\le \tau(s)}\sup_{t\in[t_k,t_{k+1}]}|\x(t)-y(t)|\le
(M_{s-1}+2D)\d.\hphantom{} \label{0d} \eaa
 For certainty, we assume that $h_{s-1}=-1$. Since $s\in\T$, it follows that
$h_s=1$ and $\x(t_{s-1})\in [y_{s},y_{s-1}]$.   If $s+1\in\T$ or
$s+2\in\T$ then (\ref{tree}) holds. Suppose that $s+1\notin\T$,
$s+2\notin\T$, and $s+3\notin\T$.   Hence \baaa \x(t_{s+3})\in
[y_{s}-4D\d,y_{s-1}+4D\d].\eaaa Since $s+1\notin\T$ and
$s+2\notin\T$, it follows that  $\x(t_{s+3})\in
[y_{s-1},y_{s-1}+4D\d)$. On the other hand, \baaa
y(t_{s+3})=y_{s-1}-M_{s-1}\d +M_s\d+M_{s+1}\d+M_{s+2}\d\brea\ge
y_{s-1}+2M_s\d\ge y_{s-1}+4D\d.\eaaa

 It follows that $\{s+1,s+2,s+3\}\cap\T\neq\emptyset$ and  (\ref{tree}) holds.

Let us prove (\ref{0d}). We have to consider the cases where $\tau(s)=s+1,s+2,s+3$  separately.
Let us assume again that $h_{s-1}=-1$ and  $h_s=1$.
\par
We have that $M_{s-1}\ge 2aD$ and $\x(t_s)\in[y_{s-1}+D\d,y_s]$.  \index{and  \baaa
&&\x(t)\in [y_{s-1}+D\d,y_{s}-D\d],\quad t\in[t_{s-1},t_{s}],\\
&&y(t)=y_{s-1}-M_{s-1}(t-s)\d,\quad t\in[t_{s-1},t_{s}].
\eaaa
Hence
 \baaa
&&|\x(t)-y(t)|\le (M_{s-1}+D)\d,\quad  t\in[t_{s-1},t_{s}].
\eaaa}
\par Let us assume that $\tau(s)=s+1$.
In this case,
 \baaa &&\x(t_{s+1})\le y_{s+1},\\
&&\x(t_{s+1})\in [y_{s}-D\d,y_{s+1}],\quad\x(t_s)\in
[y_{s},y_{s+1}+D\d],\\
&&\x(t)\in [y_{s}-D\d,y_{s+1}+D\d],\qquad t\in[t_{s},t_{s+1}],\\
&&y(t)=y_{s}+M_{s}(t-t_s)\d,\quad t\in[t_{s},t_{s+1}]. \eaaa
Hence (\ref{0d}) holds for the case where $\tau(s)=s+1$.
\par
Let us assume that $\tau(s)=s+2$. In this case,  \baaa
&&\x(t_{s+2})\le y_{s+2},\quad  \x(t_{s+1})> y_{s+1},\\
&&\x(t_{s+2})\in [y_{s+1}-D\d,y_{s+2}],\quad\breakk\x(t_{s+1})\in [y_{s+1},y_{s+2}+D\d],\\
&&\x(t_{s})\in [y_{s+1}-D\d,\min(y_{s-1}+D\d,y_{s+2}+2D],\eaaa
and\baaa
&&\x(t)\in [y_{s+1}-D\d,y_{s+2}+D\d],\quad t\in[t_{s+1},t_{s+2}],\\
&&\x(t)\in [y_{s+1}-D\d,\min(y_{t-s}+2D\d,y_{s+2}+D\d)],\breakk\quad t\in[t_{s},t_{s+1}],\\
&&y(t)=y_{s+i}+M_{s+i}(t-t_{s+i})\d,\quad
t\in[t_{s+i},t_{s+i+1}],\breakk\quad i=0,1,2. \eaaa Hence (\ref{0d})
holds for the case where $\tau(s)=s+2$.
\par
Let us assume that $\tau(s)=s+3$.
In this case,  \baaa
\x(t_{s+3})\le y_{s+3},\quad  \x(t_{s+2})> y_{s+2},\quad   \x(t_{s+1})> y_{s+1},\eaaa and
 \baaa
&&\x(t_{s+3})\in [y_{s+2}-D\d,y_{s+3}],\quad\\&&
\x(t_{s+2})\in [y_{s+2},\min(y_{s-1}+3D\d,y_{s+3}+D\d],\quad\\ &&\x(t_{s+1})\in [\max(y_{s+1},y_{s+2}-D\d),\min(y_{s-1}+2D\d,\breakk y_{s+3}+2D) ],\\&&
\quad \x(t_{s})\in [y_{s+1}-D\d,\min(y_{s-1}+D\d,y_{s+2}+2D)].\eaaa
In addition, \baaa
&&\x(t)\in [y_{s+2}-D\d,y_{s+3}+D\d],\quad t\in[t_{s+2},t_{s+3}],\\
&&\x(t)\in [y_{s+1}-D\d,y_{s-1}+3D\d],\quad t\in[t_{s+1},t_{s+2}],\\
&&\x(t)\in [y_{s+1}-D\d,y_{s+2}+D\d],\quad t\in[t_{s},t_{s+1}],\\
&&y(t)=y_{s+i}+M_{s+i}(t-t_{s+i})\d,\quad
t\in[t_{s+i},t_{s+i+1}],\breakk\quad i=0,1,2,3. \eaaa
Hence (\ref{0d}) holds for the case where $\tau(s)=s+3$.
\index{We have that \baaa
&&\x(t)\in [y_{s-1}+K(t-s)\d+D\d,y_{s}+K(t-s)\d+D\d],\\
&&y(t)\in [y_{s-1}-K(t-s)\d+D\d,y_{s-1}+K(t-s)\d+D\d],
\eaaa}
\par
Let us prove  that \baa M_{\tau(s)}\le \max(M_{s-1}/a,\oo
M).\label{min}\eaa We found above that $\rho=\tau(s)\le s+3$. By the
definitions, $M_\rho=\max(M_{\rho-1}/a,\oo M)$. Further, if
$\rho=s+1$ then $M_\rho=\max(M_{s}/a,\oo M)=\max(\max(M_{s-1}/a,\oo
M)/a,\oo M)\le \max(M_{s-1}/a,\oo M) $. If $\rho=s+2$ then
$M_{s+1}=M_s$ and $M_\rho=\max(M_{s+1}/a,\oo M)=\max(M_{s}/a,\oo
M)\le \max(M_{s-1}/a,\oo M) $, similarly to the previous case.
 If $\rho=s+3$ then $M_{s+1}=M_s=\max(M_{s-1}/a,\oo M)$, $M_{s+2}=aM_{s+1}=aM_s$, and
$M_\rho=\max(M_{s+2}/a,\oo M)=\max(M_{s+1},\oo M)=\max(M_{s},\oo M)=\max(M_{s-1}/a,\oo M)$.
      Then (\ref{min}) follows.
\par
Let us prove statement (ii).  Let us define $\t_0=\tau(0)$,
$\t_k=\tau(\t_{k-1})$, $k>0$. By (\ref{tree}) and (\ref{min}) applied
to $\t_k$ instead of $\tau(0)$,  it follows that $\t_k-\t_{k-1}\le
3$ and $M_{\t_k}=\max(M_{\t_{k-1}-1}/a,\oo M)$. Hence
$M_{\t_k}=\max(a^{-k+1}M_{\tau(0)},\oo M)$. It follows that $\eta$
exists and $\eta\le \t_k$, where $k\le \log_a(M_{\tau(0)}/\oo M)+2$
and $\t_k\le \tau(0)+3k$.

Let us prove statement (iii). Let us observe that, in the sequence
$(h_{\eta+1},h_{\eta+2},h_{\eta+3},....)$, there are no quadruple
occurrences of the same symbol, i.e., for all $m\ge \eta$,
\baa(h_{m+1},h_{m+2},h_{m+3},h_{m+4})\neq
\pm(1,1,1,1).\label{nopm}\eaa We will use the induction method.
Assume that the statement holds for $k\in[\eta,m]$, where $m\in\T$.
It suffices  to show that there exists $m_0\in\{m+1,m+2,m+3\}\cap\T$
such that the statement holds for $k\in\{m+1,...,m_0\}$.
 For certainty, we assume that $h_m=1$.  This means that $M_m=\oo M$ and $h_{m-1}=-1$. \begin{itemize}
\item
Assume that $h_{m+1}=-1$. It follows that   $M_{m+1}=\oo M$ and $m+1\in\T$.
\item
Assume that $(h_{m+1},h_{m+2})=(1,-1)$. It follows that   $(M_{m+1},M_{m+2})=(\oo M,\oo M)$
 and $m+2\in\T$.
\item
Assume that $(h_{m+1},h_{m+2})=(1,1)$. It follows from (\ref{tree}) that $h_{m+3}=-1$. Hence   $(M_{m+1},M_{m+2},M_{m+3})=(\oo M,a\oo M,\oo M)$
 and $m+3\in\T$.
\end{itemize}
In addition, (\ref{0d}) holds for $s=m$.
By induction, the proof of (iii) follows. This completes the proof of
Theorem \ref{ThM}. $\Box$ \index{\section{Acknowledgement} This work
was supported by the Australian Research Council.}

\end{document}